\documentclass{amsart}
\usepackage{amssymb}

\theoremstyle{plain}

\numberwithin{equation}{section}
\input{tcilatex}

\begin{document}
\title{Significance of Negative Energy states in Quantum Field Theory $(1)$}
\author{Shi-Hao Chen}
\address{Institute of Theoretical Physics, Northeast Normal University \\
Changchun 130024, China. ~~}
\email{shchen@nenu.edu.cn }
\date{Mar. 23, 2002}
\dedicatory{}
\thanks{}

\begin{abstract}
We suppose that there are both particles with negative energies described by 
$\mathcal{L}_{W}$ and particles with positive energies described by $%
\mathcal{L}_{F\text{ }},$ $\mathcal{L}_{W}$ and $\mathcal{L}_{F\text{ }}$
are independent of each other before quantization, dependent on each other
after quantization and symmetric, and $\mathcal{L}=\mathcal{L}_{W}$ $+$ $%
\mathcal{L}_{F\text{ }}$. From this we present a new quantization method for
QED. That the energy of the vacuum state is equal to zero is naturally
obtained. Thus we can easily determine the cosmological constant according
to data of astronomical observation, and it is possible to correct
nonperturbational methods which depend on the energy of the ground state in
quantum field theory.
\end{abstract}

\maketitle

\section{ Introduction}

There are the following five problems to satisfactorily solve in the
convertional quantum field theory (QFT).

1. The issue of the cosmological constant.

2. The problem of divergence of Feynman integrals with loop diagrams.

3. The problem of the origin of asymmetry in the electroweak\ unified theory.

4. The problem of triviality of $\varphi ^{4}-$theory.

5. The problems of dark matter and the origin of existence of huge cavities
in cosmos.

In brief, we present a consistent QFT without divergence, give a fully
method evaluating Feynman integrals (see the second and third papers), give
possible solutions to the five problems in a unified basis to reexplain the
physics meanings of negative energies.

There is an inconsistency in the conventional QFT.

As is well known, before redefining a Hamilton $H$ \ as a normal-ordered
product, the energy $E_{0}$ of the vacuum state is divergent. Because we may
arbitrarily choose the zero point of energy in QFT, we can redefine $E_{0}$
to be zero. This is, in fact, equivalent to demand 
\begin{equation*}
\{a_{\mathbf{p}s},a_{\mathbf{p}s}^{+}\}=\{b_{\mathbf{p}s},b_{\mathbf{p}%
s}^{+}\}=[c_{\mathbf{k}\lambda },c_{\mathbf{k}\lambda }^{+}]=0,
\end{equation*}%
in the conventional $QFT$. But in fact these commutation relations are equal
to 1, and in other cases, e.g. in propagators, they must also be 1. Thus the
conventional QFT is not consistent.

Divergence of Feynman integrals with loop diagrams seems to have been solved
by introducing the bare mass and the bare charge or the concepts equivalent
to them. But both bare mass and charge are divergent and unmeasured, thus
QFT is still not perfect. In order to overcome the shortcomings, people have
tried many methods. For example, G. Scharf attempted to solve the difficulty
by the causal approach $^{\left[ 1\right] }.$ ~Feynman integrals with loop
diagrams are not divergent~in some supersymmetric theories. But the
supersymmetry theory lacks experiment foundations. In fact, there should be
no divergence and all physical quantities should be measurable in a
consistent theory.

According to the given generalized electroweak unified models which are
left-right symmetric before symmetry spontaneously breaking, asymmetry is
caused by symmetry spontaneously breaking. In such models there must be many
unknown particles with massive masses. Such models are troublesome and
causes many new problems. Hence the origin of asymmetry in the electroweak\
unified theory should still be explored.

In order to introduce the present theory, we first discuss measurement of an
energy. For a measurable energy, in fact, there is the following conjecture.

\begin{description}
\item[Conjecture 1.] \textbf{\ }Any measurable energy $E$\ of a physical
system must be a difference between two energies $E_{i}^{\prime }$\ and $%
E_{j}^{\prime }$\ belonging respectively to two states $\mid $\ $%
E_{i}^{\prime }$\ $\rangle $\ and $\mid E_{j}^{\prime }\rangle ,$\ i.e., 
\begin{equation}
E=\langle E_{i}^{\prime }\mid H\mid E_{i}^{\prime }\rangle -\langle
E_{j}^{\prime }\mid H\mid E_{j}^{\prime }\rangle =E_{i}^{\prime
}-E_{j}^{\prime },  \tag{1.1}
\end{equation}%
where $\mid E_{j}^{\prime }\rangle $\ is such a state into which $\mid $\ $%
E_{i}^{\prime }$\ $\rangle $\ can transit by radiating ~the energy $~E.$
\end{description}

It is seen from the conjecture 1 that in order to determine a measurable
energy $E$ we must firstly determine two states $\mid E_{j}^{\prime }\rangle 
$ and $\mid E_{i}^{\prime }\rangle ,$ and only when $\mid E_{i}^{\prime
}\rangle $ can transit into $\mid E_{j}^{\prime }\rangle $ by radiating the
energy $\mathit{~}E,$ $E$ is measurable. Let $E_{\min }^{\prime }$ be the
minimal energy of a system$,$ an energy $E$ of the system can be defined as $%
E=E_{j}^{\prime }-E_{\min }^{\prime }.$ It is obvious that $E\geqslant 0.$ $%
E_{\min }^{\prime }=?$ $\ E_{\min }^{\prime }=0$ is necessary. According to
the general relativity 
\begin{equation}
m_{g}=m_{i}=E,  \tag{1.2}
\end{equation}%
where $m_{g}$ and $m_{i}$ are the gravitational mass and the inertial mass
corresponding to $E,$ respectively. If $E_{\min }^{\prime }\neq 0,$ $E_{\min
}^{\prime }$ will cause a gravitational effect. Thus $E_{\min }^{\prime }$
is measurable (let the cosmological constant be given). If $E_{\min
}^{\prime }$ is a measurable energy, $E_{\min }^{\prime }$ $\geqslant 0.$ If 
$E_{\min }^{\prime }$ $>0$, there must be another state $\mid E_{\min
}^{\prime \prime }\rangle $ with its energy $E_{\min }^{\prime \prime
}<E_{\min }^{\prime },$ thus $\mid E_{\min }^{\prime }$ $\rangle $ is no
longer the ground state. Hence we should have

\begin{equation}
E_{\min }^{\prime }=0.  \tag{1.3}
\end{equation}%
But before Hamiltonian operator is defined as a normal-ordered product, the
energy $E_{0}=E_{\min }^{\prime }$ is divergent, thus the convintional $QFT$
is inconsistent with the general theory of relativity and cause the issue of
cosmological constant. Of course, if the general theory of relativity is not
considered, $E_{\min }^{\prime }$ is unmeasured and undertermined.

It is seen from the above mentioned that the conventional QFT is not perfect
and should be corrected.

We introduce the present theory as follows.

1. A new Lagragian density

The physics basis of the present theory is to reexplain the physics meanings
of negative energies. The relativistic theory is very perfect, and existence
of negative energies is its essential character. Any existence must depend
on its existing conditions. As an ancient Chinese philosopher said, nothing
in nothing is just some existence; existence in existence is just some
nothing. We think that positive energies and negative energies depend on
each other, not only are negative energies not a difficulty, but have
profound physical meanings. Existence of antiparticles is only, in fact, a
result of particle-inversion symmetry, and do not reveal the essence of
negative energies. For example, a pure neutral particle or a pure neutral
world also\ has its negative-energy states. From this, we think that
existence of positive energies and negative energies imply that there are
two sorts of matter in form and the two sorts of matter are symmetric. $[2]$
supposes that the two sorts of matter possess all positive energies and try
to solve the problems above. In fact, this is not necessary. Because $%
\mathcal{L}_{F}$ and $\mathcal{L}_{W}$ \ are independent of each other,
there is no interaction and mutual transformation of the two sorts of
matter. Thus no contradiction will appear if there is matter with negative
energy. On the basis of the conjecture 1, in relativistic quantum mechanics
(RQM) frame we present the conjecture 2. In contrast with [2], from the
conjecture we will see that there are particles with negative energyies as
well as particles with positive energies.

\begin{description}
\item[Conjecture 2.] Any particle can exist in two sorts of forms ------ $%
\mid E^{+},q,1\rangle $ and $\mid E^{-},\overline{q},0\rangle $ or $\mid
E^{+},\overline{q},1\rangle $ and $\mid E^{-},q,0\rangle $, where $\mid
E^{-},q,0\rangle $ ($\mid E^{-},\overline{q},0\rangle $ )\ is a vacuum state
with the potential energy $-E^{-}$ \ and the quantum number $-q$ $(-%
\overline{q})$ \ relative to the groud state $\mid E^{-},q,1\rangle $ ($\mid
E^{-},\overline{q},1\rangle $).The two sorts of existing forms can transform
from one to another. $\mathcal{L}=\mathcal{L}_{F}+\mathcal{L}_{W}$, $%
\mathcal{L}_{F}$\ describing $\mid E^{\pm },q,1\rangle $ \ and $\mathcal{L}%
_{W}$ \ describing $\mid E^{\pm },\overline{q},1\rangle $ are independent of
each other before quantization. Particles described by\ $\mathcal{L}_{F}$\ \
($F$-particles) and the particles described by $\mathcal{L}_{W}$\ ($W$%
-particles) are symmetric.
\end{description}

We explain the conjecture 2 as follows.

The bases of the conjecture are that

A. there are always positive solutions $\mid E^{+},q,1\rangle $ and negative
solutions $\mid E^{-},q,1\rangle $ for any particle in RQM, $E^{+}$ and $%
E^{-}$ are strict symmetric in relativistic theory, and the property $%
E^{+}>0 $ \ and $E^{-}<0$ \ is relativistic invariant, where $q$ denotes the
quantum number of a particle, $E^{\pm }=\pm \sqrt{\mathbf{p}^{2}+m^{2}}$ and
1 denotes the number of particle;

B. analogously to the Dirac's hole theory, $\mid E^{-},\overline{q},0\rangle 
$ as $\mid E^{+},q,1\rangle $ is regarded as an existing state with $E^{+}$
and $q;$

C. any particle is the same in existing form. According to the Dirac's hole
theory, $e^{+}$ and $e^{-}$ are different in existing form, i.e.,$e^{+}$
exists as a hole and $e^{-}$ exists as a particle. According to the
conjecture, $e^{+}$ exists in $\mid E^{-},e^{-},0\rangle $ or $\mid
E^{+},e^{+},1\rangle ,$ and $e^{-}$ exists in $\mid E^{-},e^{+},0\rangle $
or $\mid E^{+},e^{-},1\rangle ,$i.e., $e^{+}$ and $e^{-}$ are the same in
existing form.

The physical meanings of $E^{+}$ and $E^{-}$ are different in essence. If
positive energies and negative energies are respectively conservational, $%
\{\mid E^{+},q,1\rangle \}$ and $\{\mid E^{-},q,1\rangle \}$ will correspond
to two different sorts of particles. Of course, positive energies and
negative energies are not respectively conservational in RQM frame. But
according to the conjecture 1, we will see that any particle exists in two
sorts of forms.

Because only when there is $\mid E^{-},q,0\rangle $, can a particle $q$
transit to the states with $E^{-}$ and $q$ \ and $\mid E^{-},q,1\rangle $
can come into being. Hence when there is $\mid E^{-},q,0\rangle ,$the ground
state of the system is not $\mid 0\rangle $, but $\mid E^{-},q,1\rangle .$ $%
\mid E^{-},q,0\rangle $ \ is different from $\mid 0\rangle $. When $\mid
E^{-},q,0\rangle $ \ becomes the final state $\mid E^{-},q,1\rangle $ \ by
some way, according to the conjecture 1 the energy and the additive quantum
number released by the system are respectively \ 
\begin{equation}
E^{-}\cdot 0-E^{-}\cdot 1=-E^{-}=E^{+},  \tag{1.4}
\end{equation}%
\begin{equation}
q\cdot 0-q\cdot 1=-q\equiv \overline{q}.  \tag{1.5}
\end{equation}%
Let there be $\mid E^{+},q,1\rangle $ whose ground state is $\mid E^{\prime
},q^{\prime },n\rangle $ \ (e.g., $\mid E^{\prime },q^{\prime },n\rangle
=\mid 0\rangle ,$ $\mid E^{-},\overline{q},1\rangle $ \ etc..). When $\mid
E^{+},q,1\rangle $ \ becomes the final state $\mid E^{\prime },q^{\prime
},n\rangle $, the energy and the additive quantum number released by the
system are respectively 
\begin{equation}
(E^{+}\cdot 1+nE^{\prime })-nE^{\prime }=E^{+},  \tag{1.6}
\end{equation}%
\begin{equation}
(q\cdot 1+nq^{\prime })-nq^{\prime }=q.  \tag{1.7}
\end{equation}%
It is seen that like existence of $\mid E^{-},q,1\rangle ,$existence of $%
\mid E^{-},q,0\rangle $ is not unconditional, but conditional, i.e., $\mid
E^{-},q,0\rangle $ \ should be regarded as an excited state.$\mid
E^{-},q,0\rangle $ is equivalent to $\mid E^{+},\overline{q},1\rangle $ $\ $%
with the same ground state $\mid E^{-},q,1\rangle $. But the existing forms
of $\mid E^{+},q,1\rangle $ and $\mid E^{-},q,0\rangle $ \ are different. $%
\mid E^{+},q,1\rangle $ \ exists as a particle, and $\mid E^{-},q,0\rangle $
\ exists as `nothing in nothing', and is analogous to a hole in Dirac's
theory. It should be emphasized that because the number of particles $n=0,$
the properties (energy and quantum number) of the $\mid E^{-},q,0\rangle $ \
cannot directly be determined by Lagrangian density, but can only be
determined by the particle state which $\mid E^{-},q,0\rangle $ \ transforms
into ( i.e., $\mid E^{-},q,1\rangle $ \ and $\mid E^{+},\overline{q}%
,1\rangle ,$ see below). We regard $\mid E^{-},q,0\rangle $ \ whose ground
state is $\mid E^{-},q,1\rangle $ as such a vacuum state with potential
energy $-E^{-}$ \ and quantum number $-q.$ $\mid E^{-},q,0\rangle $ \ is a
vacuum state different from $\mid 0\rangle $ \ whose ground state is still $%
\mid 0\rangle $. We cannot measure the potential,but can only measure an
energy of a particle. We suppose that $\mid E^{-},q,0\rangle $ \ and a
particle state can transform from one to another by the following process.%
\begin{equation}
\mid E^{-},q,0\rangle \rightleftarrows \mid E^{-},q,1\rangle \mid E^{+},%
\overline{q},1\rangle .  \tag{1.8}
\end{equation}%
The ground state of the right-side in (1.8) is still $\mid E^{-},q,1\rangle
, $ hence from (1.1) we obtain the energy and quantum number of the
right-side to be respectively%
\begin{equation}
(E^{-}+E^{+})-E^{-}=-E^{-}=E^{+},  \tag{1.9}
\end{equation}%
\begin{equation}
(q+\overline{q})-q=\overline{q}.  \tag{1.10}
\end{equation}%
The process does not destroy any conservation law, hence it may occur.

In RQM\ frame, $\mid E^{+},\overline{q},1\rangle $ \ cannot be described by $%
\mathcal{L}_{F}$ describing $\{\mid E^{\pm },q,1\rangle \},$ a Lagrangian
density describing $\{\mid E^{+},\overline{q},1\rangle \}$ must be different
from $\mathcal{L}_{F},$ and it is denoted by $\mathcal{L}_{W}.$ Because $%
E^{+}$ and $E^{-}$ are strict symmetric in relativistic theory, $\mid
E^{+},q,1\rangle $ \ and $\mid E^{-},q,1\rangle $ \ must be symmetric,$\mid
E^{-},q,1\rangle $ \ and $\mid E^{+},\overline{q},1\rangle $ \ are also
symmetric, hence $\mid E^{+},\overline{q},1\rangle $ $\ $and $\mid
E^{+},q,1\rangle $ must be symmetric. Thus we suppose that $\mathcal{L}_{F}$%
\textbf{\ }and\textbf{\ }$\mathcal{L}_{W}$ \ are symmetic, independent of
each other before quantization and $\mathcal{L=L}_{F}+\mathcal{L}_{W.}$
Because $\mathcal{L}_{F}$\textbf{\ }and\textbf{\ }$\mathcal{L}_{W}$ \ are
symmetic, there are states $\{\mid E^{\pm },\overline{q},1\rangle \}$ \ of
the particle $\overline{q}$ \ and the process

\begin{equation}
\mid E^{-},\overline{q},0\rangle \rightleftarrows \mid E^{-},\overline{q}%
,1\rangle \mid E^{+},q,1\rangle .  \tag{1.11}
\end{equation}%
Thus, both $q$ and $\overline{q}$ have two sorts of existing forms, i.e., $q$
exists in forms $\mid E^{+},q,1\rangle $ and $\mid E^{-},\overline{q}%
,0\rangle $ \ (their grounds are all $\mid E^{-},\overline{q},1\rangle $ ) , 
$\overline{q}$ exists in forms $\mid E^{+},\overline{q},1\rangle $ and $\mid
E^{-},q,0\rangle $ \ (their grounds are all $\mid E^{-},q,1\rangle $ ), and $%
q$ and $\overline{q}$ are symmetric and the two sorts of existing forms can
transform from one to another.

From the conjecture 1 we see that only when all physics quantities of the
ground state of a physics system are invariant, is it convenient to discuss
evolution of the system, and we can compare physics quantities of two states
or two systems. Hence we present the following conjecture for the ground
state of a system.

\begin{description}
\item[Conjecture 3.] All physics quanties of the ground state of a system
are always invariant.
\end{description}

According to the conjecture, there are processes analogous to the following
process 
\begin{equation}
\mid E_{1}^{-},e^{-},0\rangle +\mid E_{2}^{-},e^{-},1\rangle
\leftrightharpoons \mid E_{1}^{+}+E_{2}^{+},\gamma ,1\rangle +\mid
E_{1}^{-},e^{-},1\rangle ,  \tag{1.12}
\end{equation}%
\ but there are not such processes analogous to 
\begin{equation}
\mid E_{1}^{+}+E_{2}^{+},\gamma ,1\rangle +\mid E_{1}^{-},e^{-},1\rangle
\rightarrow \mid E_{2}^{+},e^{-},1\rangle .  \tag{1.13}
\end{equation}%
We will prove in another paper that in RQM\ frame from conjectures 2 and 3
we can obtain the same results as those obtained from conventional RQM.

In contrast with the symmetry of a particle and its antiparticle, the
symmetry is a strict symmetric which is not destroyed in any interaction, on
the other hand, even $q$ is a pure neutral particle, i.e., $q=\overline{q},$ 
$\mid E^{+},\overline{q},1\rangle $ \ and $\mid E^{+},q,1\rangle $ \ cannot
be regarded as the same particle since $q$ is in $\mathcal{L}_{F}$ , $%
\overline{q}$ is in $\mathcal{L}_{W}$ \ and $\mathcal{L}_{F}$\textbf{\ }and%
\textbf{\ }$\mathcal{L}_{W}$ \ are independent of each other. $q$ in $%
\mathcal{L}_{F}$ \ and $\overline{q}$ in $\mathcal{L}_{W}$ \ are two
freedoms. In contrast with the Dirac's hole theory in which the electron sea
with the infinite negative energy is necessary, in the present theory, $%
e^{+} $ and $e^{-}$ are the same in existing form and the electron sea is no
longer necessary.

Of course, if only in RQM frame to discuss matter, $\mathcal{L}_{W}$ \ is
unnecessary since we cannot obtain new measuable results by $\mathcal{L}_{F}+%
\mathcal{L}_{W}$. But after quantization, $\mathcal{L}_{F}$\ and $\mathcal{L}%
_{W}$ will be dependent on each other, and the present theory is different
from the conventional QFT. The discussion above and the conjecture 1-3 may
be regarded as a transition.

In QFT frame, fields become field operators. We will see that energies $%
E^{+} $ determined by $\mathcal{L}_{F}$ \ are all positive, energies $E^{-}$
determined by $\mathcal{L}_{W}$ \ are all negative, and it is unnecessary to
consider states analogous to $\mid E^{-},q,0\rangle $. Thus, in QFT the
ground state of the world described by $\mathcal{L}_{F}$ is still the vaccum
state $\mid 0\rangle $, the state with the higuest energy of the world
described by $\mathcal{L}_{W}$ is also the vaccum state $\mid 0\rangle .$
Because there is no coupling between the fields in $\mathcal{L}_{F}$ \ and
the fields in $\mathcal{L}_{W}$, the energy $E^{+}$ \ determined by $%
\mathcal{L}_{F}$ \ and the energy $E^{-}$ determined by $\mathcal{L}_{W}$ \
are respectively conservational, and a F-particle and a W-particle cannot
transform from one to another in fact by interaction determined by $\mathcal{%
L}$. Thus, in QFT the conjectures 1 and 3 are not necessary, and the
conjecture 2 become the following form.

\begin{description}
\item[Conjecture A.] Any particle can exist in two sorts of states ------$%
F-\mid E^{+},q,1\rangle $ described by $\mathcal{L}_{F}$\ and $W-\mid
E^{+},q,1\rangle $ described by $\mathcal{L}_{W}$. $\mathcal{L=L}_{F}+%
\mathcal{L}_{W}$ , $\mathcal{L}_{F}$\ and $\mathcal{L}_{W}$ are independent
of each other before quantization and dependent on each other after
quantization. The particles described by\ $\mathcal{L}_{F}$\ and the
particles described by $\mathcal{L}_{W}$\ are symmetric.
\end{description}

According to the conjecture, every particle in $\mathcal{L}_{F}$ is
accordant with a particle in $\mathcal{L}_{W}$, and the properties of the
two particles are the same, e.g., there are two sorts of electrons, i.e., a $%
F-$electron with a positive energy and a $W-$electron with a negative
energy. That $\mathcal{L}_{F}$ and $\mathcal{L}_{W}$ are independent of each
other implies that there is no coupling between fields in $\mathcal{L}_{F}$
and fields in $\mathcal{L}_{W},$ but after quantization, $\mathcal{L}_{F}$
and $\mathcal{L}_{W}$ \ will be dependent on each other. Thus, the two sorts
of energies corresponding to $\mathcal{L}_{F}$ and $\mathcal{L}_{W}$ are
respectively conservational, a real particle in $\mathcal{L}_{F}$ cannot
transform into a real particle in $\mathcal{L}_{W}$, and vice versa. But the
two sorts of virtual particles can transform from one to another.

We call the conjecture A the F-W (fire-water) symmetry conjecture. We may
also call conjecture A the L-R (left-right) symmetry conjecture since $%
\mathcal{L}_{F}$\ and $\mathcal{L}_{W}$ describe respectively the left-hand
world (matter world) and right-hand world (dark-matter world) and $\mathcal{L%
}_{F}\ +\mathcal{L}_{W}$ is left-right symmetry (see paper 3), or $%
E^{+}-E^{-}$ symmetry conjecture since $E^{\pm }$ is the basis of the
conjecture.

From conjecture A we can obtain possible solutions to the five problems
above.

2. Transformation operators and a new method quantizing fields.

Because particles can exist in the two sorts of forms, we can define
transformation operators which transform a F-particle into a W-particle or a
W-particle into a F-particle, and can quantize fields by the transformation
operators replacing creation and annihilation operators in the conventional
QFT. Thus it is necessary that $g_{f}$ and $m_{ef\text{ }}$ respectively
become operators $G_{F}$ \ and $M_{F}$ \ to be determined by $\mathcal{S}%
_{w} $ ,\ and $g_{w\text{ \ }}$and $m_{ew\text{ }}$ respectively become
operators $G_{W}$ \ and $M_{W}$ to be determined by $\mathcal{S}_{f}$, here $%
\mathcal{S}_{w}$ and $\mathcal{S}_{f}$ are the scattering operators
respectively determined by $\mathcal{L}_{W}$ and $\mathcal{L}_{F}$. $G_{F}$
\ and $M_{F}$ multiplied by field operators $\psi $ \ and $A_{\mu }$\ become
the coupling coefficient $g_{f}(p_{2},p_{1})$ \ and mass $m_{ef\text{ }}(p)$
determined by scattering amplitude $\langle W_{f}\mid S_{w}\mid W_{i}$ $%
\rangle ,$ and $G_{W}$ \ and $M_{W}$ multiplied by field operators $%
\underline{\psi }$ \ and $\underline{A}_{\mu }$ \ become $g_{w}(p_{2},p_{1})$
\ and $m_{ew\text{ }}(p)$ determined by scattering amplitude $\langle
F_{f}\mid S_{f}\mid F_{i}$ $\rangle $\ . Thus after quantization, $\mathcal{L%
}_{F}$ \ and $\mathcal{L}_{W}$ will be dependent on each other.

3. Two sorts of corretions.

In the conventional QED, there are two sort of parameters, e.g., the
physical charge and the bare charge, and one sort of corrections originating 
$\mathcal{S}$ \ equivalent to $\mathcal{S}_{f}.$ In contrast with the given
QED, there are only one sort of parameters defined at so-called subtraction
point $q_{2},$ $q_{1}$ and $q\prime ,$ i.e., $%
g_{f}(q_{2},q_{1})=g_{w}(q_{2},q_{1})=g_{0}$ \ and $%
m_{ef}(q)=m_{ew}(q)=m_{e0},$ and two sorts of corretions originating from $%
\mathcal{S}_{w}$ and $\mathcal{S}_{f}$ to scattering amplitudes, $g_{0}$ \
and $m_{e0}.$Thus $\mathcal{L}_{F}$ and $\mathcal{L}_{W}$ together determine
the loop-diagram corrections. When n-loop corrections originating from $%
\mathcal{S}_{f}$ and $\mathcal{S}_{w}$ \ are simultaneously considered, the
integrands causing divergence in $\langle F_{f}\mid S_{f}\mid F_{i}$ $%
\rangle $ or $\langle W_{f}\mid S_{w}\mid W_{i}$ $\rangle $ will cancil each
other out, hence all Feynman integrals are convergent, e.g.,%
\begin{equation*}
g_{f}^{(1)}(p_{2},p_{1})=g_{ff}^{(1)}(p_{2},p_{1})+g_{fw}^{(1)}(q_{2},q_{1})
\end{equation*}%
is finite and $g_{f}^{(1)}(q_{2},q_{1})=0$, where $g_{ff}^{(1)}$ \
originates from $\mathcal{S}_{f}$ \ and $g_{fw}^{(1)}$ \ originates from $%
\mathcal{S}_{w}$, and the superscript (1) denotes 1-loop correction$.$ Thus
it is unnecessary to introduce counterterms and regularization. We give a
complete Feynman rules to evaluate Feynman integrals by the new concepts
(see second and third papers).

It should be pointed that\ in the meaning of perturbation theory, because
the coupling coefficients and masses\ will be corrected by n-loop diagrams,\
we cannot give a absolutely precise $\mathcal{L}_{F}$ and $\mathcal{L}_{W}$
\ in the prime, and can only give the precise $\mathcal{L}_{F}^{(0-loop)}$
and $\mathcal{L}_{W}^{(0-loop)}$\ at the subtraction point or approximate to
tree diagrams. Of course, by such $\mathcal{L}_{F}^{(0-loop)}$ and $\mathcal{%
L}_{W}^{(0-loop)}$ we can obtain scattering amplitudes approximate arbitrary
n-loop diagrams.

4. $\langle 0\mid H\mid 0\rangle \equiv E_{0}=E_{0F}+E_{0W}=0$ is naturally
derived, thereby we can easily determine the cosmological constant according
to data of astronomical observation, and it is possible to correct
nonperturbational methods which depend on the energy of the ground state in
QFT.

When $\psi $ and $A_{\mu }$ $\ $in $\mathcal{L}_{F}$ \ and \ $\underline{%
\psi }$ \ and $\underline{A}_{\mu }$ \ in $\mathcal{L}_{W}$ are regarded as
the classical fields or the coupling coefficients $g_{f\text{ \ }}$and the
electromagnetic masses $m_{ef\text{ }}$in $\mathcal{L}_{F}$ \ , $g_{w\text{
\ }}$and $m_{ew\text{ }}$ in $\mathcal{L}_{W}$ are regarded as constants, $%
\mathcal{L}_{F}$ \ and $\mathcal{L}_{W}$ \ will be independent of each
other. In this case, except $E_{0}=0$, all results obtained by the present
theory will be the same as those obtained by the conventional theory.

5. Generalizing the present theory to the electroweak unified theory, we
will see a possible origin of symmetry breaking. Accoding to this model, the
world is symmetric on principle (i.e., $\mathcal{L=L}_{W}$ $+$ $\mathcal{L}%
_{F}$ is symmetric), but the world observed by us is asymmetric since $%
\mathcal{L}_{W}$ or $\mathcal{L}_{F}$ is asymmetric. In this model there is
no unknown particle with a massive mass (see the third paper).

6. Because there is no interaction between the two sorts of matter by a
given quantizable field. Only possibility is that there is repulsion or
gravitation of the two sorts of matter. Because the sort of matter described
by $\mathcal{L}_{W}$ is one new sort of matter, it is impossible from theory
to determine that there is what sort of interaction. We can only determine
the new sort of interaction from data of astronomical observation. If the
new interaction is repulsion, it is possible that the new interaction is the
reason for cosmos expansion. If the new interaction is gravitation, it is
possible that the new\ sort\ of matter is the candidate for dark matter. It
is also possible that there is new and more important relationship between
the two sorts of matter.

7. The new $QFT$ \ will also give a possible solution for the problem of
triviality of $\varphi ^{4}-$theory.

8. Two sorts of transformation.

Accoding to the present theory, there are two sorts of transformation.

The first sort of transformation must correspond to a coupling term of field
operators, e.g., $ig\overline{\psi }\gamma _{\mu }A_{\mu }\psi $ determines
the transformation $e^{+}+e^{-}\rightarrow \gamma +\gamma .$ The sort of
transformation is measurable.

The second sort of transformation is defined by the transformation operators
as $\mid \underline{a}_{\mathbf{p}s}\rangle \eqslantless a_{\mathbf{p}s}\mid 
$ $\ $and cannot correspond to any coupling term of quantizable fields $.$
The processes determined by the sort of transformation, e.g., 
\begin{equation*}
\mid \underline{a}_{\mathbf{p}s}\rangle \eqslantless a_{\mathbf{p}s}\mid
\mid a_{\mathbf{p}s}\rangle =\mid \underline{a}_{\mathbf{p}s}\rangle ,
\end{equation*}%
cannot be measured. The sort of transformation can only be potential and is
realized by the virtual-particle processes. Existing reasons of the sort of
transformation are that by it we can eliminate divergence of $E_{0}$ and
Feynman integrals with loop diagrams, explain the left-right asymmetry and
some phenomena of the cosmos, and so on.

By the conventional creation and annihilation operators in the given QFT we
can also obtain the similar results, provided we suppose $\mathcal{L}=$ $%
\mathcal{L}_{F}$ $+\mathcal{L}_{W}$ and that $g_{f}$ and $m_{f}$ are
determined by $\mathcal{S}_{w}$ $\ $and $g_{w}$ and $m_{w}$ are determined
by $\mathcal{S}_{f}$ (of course, in this case this conjecture is not
natural). It is also possible to obtain the same results but that both
F-particles and W-particles possess positive energies, provided $\mathcal{L}%
_{F}$ and $\mathcal{L}_{W}$ \ are independent of each other before
quantization.

The present theory contains three parts. The first part takes QED as example
to illuminate the method to reconstruct QFT, and give the solutions of the
issue of the cosmological constant and the problem of divergent Feynman
integrals in QED. The first part is the present paper which is composed of
the part and the following two parts. The second part discusses the problem
of triviality of $\varphi ^{4}-$theory. The third part discusses the problem
of the origin of asymmetry in the electroweak\ unified theory in detail.

The present theory contains three parts. The first part takes QED as example
to illuminate the method to reconstruct QFT, and give the solutions of the
issue of the cosmological constant and the problem of divergent Feynman
integrals in QED. The part is composed of the present paper and the
following two papers. The second part discusses the problem of triviality of 
$\varphi ^{4}-$theory. The third part discusses the problem of the origin of
asymmetry in the electroweak\ unified theory.

Quantization for free fields will be discussed in the present paper. The
outline of this paper is as follows. In section 2, we construct the
Lagrangian density of free fields. In section 3, quantization for free
fields is carried out. In section 4, we discuss the meanings of $E_{0}=0.$
Section 5 is a conclusion.

\section{Lagrangian density and equations of motion for free fields}

There must be positive-energy resolutions and negative-energy resolutions
for a classical relativistic equation of motion, and positive-energy
solutions and negative-energy solutions are symmetric. On the other hand,we
think that only equations of motion are insufficient in order to determine
the complete properties of a relativistic system. A complete Lagrangian
density is very necessary .From this, we present the following conjecture.

Conjecture:\textit{\ There are both particles with} \textit{negative
energies described by }$\mathcal{L}_{W}$ (W-particle) \textit{and particles
with positive energies described by }$\mathcal{L}_{F}$ (F-particle)\textit{,
the total Lagrangian density is}%
\begin{equation}
\mathcal{L}_{0}=\mathcal{L}_{F0}+\mathcal{L}_{W0},  \tag{2.1}
\end{equation}%
\textit{\ }$\mathcal{L}_{F0}$ and $\mathcal{L}_{W0}$ \textit{are independent
of each other and symmetric. }

Field operators are composed of transformation operators which transform a
F-particle into a W-particle or a W-particle into a F-particle. Such
transformation may occur for virtual particles. But because $\mathcal{L}%
_{F0} $ and $\mathcal{L}_{W0}$ are independent of each other,\textit{\ }such
transformation cannot occur for realistic particles.

We suppose the Lagrangian densities $\mathcal{L}_{F0}$ and $\mathcal{L}_{W0}$
for the free Dirac fields and the Maxwell fields to be respectively

\begin{equation}
\mathcal{L}_{F0}=-\overline{\psi }_{0}(x)(\gamma _{\mu }\partial _{\mu
}+m)\psi _{0}(x)-\frac{1}{2}\partial _{\mu }A_{0\nu }\partial _{\mu }A_{0\nu
},  \tag{2.2}
\end{equation}

\begin{equation}
\mathcal{L}_{W0}=\overline{\underline{\psi }}_{0}(x)(\gamma _{\mu }\partial
_{\mu }+m)\underline{\psi }_{0}(x)+\frac{1}{2}\partial _{\mu }\underline{A}%
_{0\nu }\partial _{\mu }\underline{A}_{0\nu }.  \tag{2.3}
\end{equation}%
(2.2) and (2.3) imply the Lorentz gauge to be already fixed . The difference
between (2.1) and the corresponding Lagrangian density in the given QED is $%
\mathcal{L}_{W0}$ which describes motion of particles existing in the other
form. We call $\psi $, $A_{\mu ,}$ $\underline{\psi }$ and $\underline{A}%
_{\mu }$ the F-electron field, the F-photon field, and the W-electron field,
the W-photon field, respectively. The conjugate fields corresponding to them
are respectively

\begin{equation}
\pi _{0\psi }=\frac{\partial \mathcal{L}_{0}}{\partial \dot{\psi}_{0}}=i\psi
_{0}^{+},  \tag{2.4}
\end{equation}

\begin{equation}
\pi _{0\mu }=\frac{\partial \mathcal{L}_{0}}{\partial \dot{A}_{0\mu }}=\dot{A%
}_{0\mu },  \tag{2.5}
\end{equation}

\begin{equation}
\pi _{0\underline{\psi }}=\frac{\partial \mathcal{L}_{0}}{\partial 
\underline{\dot{\psi}}_{0}}=-i\underline{\psi }_{0}^{+},  \tag{2.6}
\end{equation}

\begin{equation}
\underline{\pi }_{0\mu }=\frac{\partial \mathcal{L}_{0}}{\partial \underline{%
\dot{A}}_{0\mu }}=-\underline{\dot{A}}_{0\mu }.  \tag{2.7}
\end{equation}

From the Noether$^{\prime }$s theorem, we have

\begin{equation}
H_0=H_{F0}+H_{W0},  \tag{2.8}
\end{equation}

\begin{equation}
H_{F0}=\int d^{3}x\{\psi _{0}^{+}\gamma _{4}\left( \gamma _{j}\partial
_{j}+m\right) \psi _{0}+\frac{1}{2}(\overset{\cdot }{A}_{0\mu }\overset{%
\cdot }{A}_{0\mu }+\partial _{j}A_{0\nu }\partial _{j}A_{0\nu })\}\ , 
\tag{2.9}
\end{equation}

\begin{equation}
H_{W0}=-\int d^{3}x\{(\underline{\psi }_{0}^{+}\gamma _{4}(\gamma
_{j}\partial _{j}+m)\underline{\psi }_{0}+\frac{1}{2}(\underline{\overset{.}{%
A}}_{0\mu }\underline{\overset{.}{A}}_{0\mu }+\partial _{j}\underline{A}%
_{0\nu }\partial _{j}\underline{A}_{0\nu })\},  \tag{2.10}
\end{equation}

\begin{equation}
Q=Q_{F}+Q_{W},  \tag{2.11}
\end{equation}

\begin{equation}
Q_{F}=\int d^{3}xi\psi _{0}^{+}\psi _{0},  \tag{2.12}
\end{equation}

\begin{equation}
Q_{W}=-\int d^{3}x\underline{\psi }_{0}^{+}\underline{\psi }_{0},  \tag{2.13}
\end{equation}%
where $j=1,2,3,$ a repeated index implies summation over all values of the
index. The Euler-Lagrange equations of motion derived from Hamilton$^{,}$s
variational principle are

\begin{equation}
i\frac{\partial }{\partial t}\psi _{0}=\overset{\wedge }{H}_{0}\psi _{0}, 
\tag{2.14}
\end{equation}

\begin{equation}
\square A_{0\mu }=0,  \tag{2.15}
\end{equation}

\begin{equation}
i\frac{\partial }{\partial t}\underline{\psi }_{0}=\overset{\wedge }{H}_{0}%
\underline{\psi }_{0},  \tag{2.16}
\end{equation}

\begin{equation}
\square \underline{A}_{0\mu }=0,  \tag{2.17}
\end{equation}%
where $\overset{\wedge }{H}_{0}=\gamma _{4}\left( \gamma _{j}\partial
_{j}+m\right) .$ It is seen from (2.14)-(2.17),(2.9),(2.10), (2.12), and
(2.13) that although $\mathcal{L}_{F0}\neq \mathcal{L}_{W0},Q_{F}\neq Q_{w}$
and $H_{0F}\neq H_{0W,}$ the equations satisfied by $\underline{\psi }_{0}$
and $\underline{A}_{0\mu }$ are the same as those satisfied by $\psi _{0}$
and $A_{0\mu },$ respectively. This implies that for a relativistic physical
system, only equations of motion are insufficient for corrective description
of all properties of the system. A complete Lagrangian density is very
necessary.

When $\psi _{0}$, etc., are regarded as the classical fields and%
\begin{equation}
\partial _{\mu }A_{0\mu }=\partial _{\mu }\underline{A}_{0\mu }=0, 
\tag{2.18}
\end{equation}%
$\psi $ and $\underline{\psi }$ can be expanded in terms of the complete set
of plane-wave solutions

\begin{equation}
\frac{1}{\sqrt{V}}u_{\mathbf{p}s}e^{ipx},\qquad \frac{1}{\sqrt{V}}v_{\mathbf{%
p}s}e^{-ipx},\quad \text{ \hspace{0in}}s=1,2,  \tag{2.19}
\end{equation}%
where $px=\mathbf{px}-E_{\mathbf{p}}t,$ $E_{\mathbf{p}}=\sqrt{\mathbf{p}%
^{2}+m^{2}},$ and the complete set of plane-wave solutions to (2.15) and
(2.17) is

\begin{equation}
\frac{1}{\sqrt{2\omega _{\mathbf{k}}V}}e_{\mathbf{k}\mu }^{\lambda }e^{\pm
ikx},  \tag{2.20}
\end{equation}%
where $kx=\mathbf{kx}-\omega _{\mathbf{k}}t,\omega _{\mathbf{k}}=\mid 
\mathbf{k}\mid $ , $\lambda =1,2,.$ To get a completeness relation, it is
necessary to form a guartet of orthonormal 4-vectors$^{\left[ 3\right] }.$

\begin{eqnarray}
e_{k}^{1} &=&\left( \mathbf{\varepsilon }_{\mathbf{k}}^{1},0\right)
,e_{k}^{2}=\left( \mathbf{\varepsilon }_{\mathbf{k}}^{2},0\right)
,e_{k}^{3}=-\left[ k+\eta \left( k\eta \right) \right] \diagup k\eta , 
\notag \\
\eta &=&\left( 0,0,0,i\right) ,e_{k}^{4}=i\eta ,\mathbf{\varepsilon }_{%
\mathbf{k}}^{1,2}\cdot \mathbf{k=0.}  \TCItag{2.21}
\end{eqnarray}%
Moreover, all four vectors are normalized to 1, i.e.,

\begin{equation*}
e_{\mathbf{k}}^{\lambda }e_{\mathbf{k}}^{\lambda ^{^{\prime }}}=\delta
_{\lambda \lambda ^{^{\prime }}},\text{ \qquad }\sum_{\lambda =1}^{4}e_{%
\mathbf{k}\mu }^{\lambda }e_{\mathbf{k}\nu }^{\lambda }=\delta _{\mu \nu .}
\end{equation*}

\section{Quantization for free fields}

We now regard $\psi _{0}$ etc. as field operators. $\psi _{0}$, $A_{0\mu ,}$ 
$\underline{\psi }_{0}$\hspace{0in}and $\hspace{0in}\underline{A}_{0\mu }$
as the solutions of the equations of the quantum fields $\left( 2.14\right)
-\left( 2.17\right) $ can also be expanded in terms of the complete sets $%
\left( 2.19\right) $ and $\left( 2.20\right) ,$ respectively, only the
expanding coefficients are all operators. Thus we have

\begin{equation}
\psi _{0}\left( x\right) =\frac{1}{\sqrt{V}}\sum_{\mathbf{p}s}\{\mid 
\underline{a}_{\mathbf{p}s}\rangle \xi _{\mathbf{p}s}\eqslantless a_{\mathbf{%
p}s}(t)\mid u_{\mathbf{p}s}e^{i\mathbf{px}}+\mid b_{\mathbf{p}%
s}(t)\eqslantgtr d\eta _{\mathbf{p}s}^{+}\langle \underline{b}_{\mathbf{p}%
s}\mid v_{\mathbf{p}s}e^{-i\mathbf{px}}\},  \tag{3.1}
\end{equation}

\begin{equation}
A_{0\mu }\left( x\right) =\frac{1}{\sqrt{V}}\sum_{\mathbf{k}}\frac{1}{\sqrt{%
2\omega _{\mathbf{k}}}}\sum_{\lambda =1}^{4}e_{\mathbf{k}\mu }^{\lambda }\{%
\underline{j}_{\mathbf{k}}\eqslantless c_{\mathbf{k}\lambda }(t)\mid e^{i%
\mathbf{kx}}+\mid \overline{c}_{\mathbf{k}\lambda }(t)\eqslantgtr \underline{%
j}_{\mathbf{(-k)}}e^{-i\mathbf{kx}}\},  \tag{3.2}
\end{equation}%
\begin{equation}
\underline{\psi }_{0}\left( x\right) =\frac{1}{\sqrt{V}}\sum_{\mathbf{p}%
s}\{\mid b_{\mathbf{p}s}\rangle \eta _{\mathbf{p}s}\eqslantless \underline{b}%
_{\mathbf{p}s}(t)\mid u_{\mathbf{p}s}e^{i\mathbf{px}}+\mid \underline{a}_{%
\mathbf{p}s}(t)\eqslantgtr d\xi _{\mathbf{p}s}^{+}\langle a_{\mathbf{p}%
s}\mid v_{\mathbf{p}s}e^{-i\mathbf{px}}\},  \tag{3.3}
\end{equation}

\begin{equation}
\underline{A}_{0\mu }\left( x\right) =\frac{1}{\sqrt{V}}\sum_{\mathbf{k}}%
\frac{1}{\sqrt{2\omega _{\mathbf{k}}}}\sum_{\lambda =1}^{4}e_{\mathbf{k}\mu
}^{\lambda }\{j_{\mathbf{k}}\eqslantless \underline{c}_{\mathbf{k}\lambda
}(t)\mid e^{i\mathbf{kx}}+\mid \underline{\overline{c}}_{\mathbf{k}\lambda
}(t)\eqslantgtr j_{(-\mathbf{k)}}e^{-i\mathbf{kx}}\},  \tag{3.4}
\end{equation}

\begin{equation}
\pi _{0\psi }=i\psi _{0}^{+}\left( x\right) \equiv \frac{i}{\sqrt{V}}\sum_{%
\mathbf{p}s}\{\mid a_{\mathbf{p}s}(t)\eqslantgtr d\xi _{\mathbf{p}s}\langle 
\underline{a}_{\mathbf{p}s}\mid u_{\mathbf{p}s}^{+}e^{-i\mathbf{px}}+\mid 
\underline{b}_{\mathbf{p}s}\rangle \eta _{\mathbf{p}s}^{+}\eqslantless b_{%
\mathbf{p}s}(t)\mid v_{\mathbf{p}s}^{+}e^{i\mathbf{px}}\},  \tag{3.5}
\end{equation}

\begin{equation}
\pi _{0\mu }=\dot{A}_{0\mu }\left( x\right) \equiv \frac{-i}{\sqrt{V}}\sum_{%
\mathbf{k}}\sqrt{\frac{\omega _{\mathbf{k}}}{2}}\sum_{\lambda =4}^{4}e_{%
\mathbf{k}\mu }^{\lambda }\{\underline{j}_{\mathbf{k}}\eqslantless c_{%
\mathbf{k}\lambda }(t)\mid e^{i\mathbf{kx}}-\mid \overline{c}_{\mathbf{k}%
\lambda }(t)\eqslantgtr \underline{j}_{(-\mathbf{k)}}e^{-i\mathbf{kx}}\}, 
\tag{3.6}
\end{equation}%
\begin{eqnarray}
\underline{\pi }_{0\psi } &=&-i\underline{\psi }_{0}^{+}(x)\equiv \frac{-i}{%
\sqrt{V}}\sum_{\mathbf{p}s}\{\mid \underline{b}_{\mathbf{p}s}(t)\eqslantgtr
d\eta _{\mathbf{p}s}\langle b_{\mathbf{p}s}\mid u_{\mathbf{p}s}^{+}e^{-i%
\mathbf{px}}  \notag \\
+ &\mid &a_{\mathbf{p}s}\rangle \xi _{\mathbf{p}s}^{+}\eqslantless 
\underline{a}_{\mathbf{p}s}(t)\mid v_{\mathbf{p}s}^{+}e^{i\mathbf{px}}\}, 
\TCItag{3.7}
\end{eqnarray}

\begin{eqnarray}
\underline{\pi }_{0\mu } &=&-\underline{\dot{A}}_{0\mu }\left( x\right)
\equiv \frac{i}{\sqrt{V}}\sum_{\mathbf{k}}\sqrt{\frac{\omega _{\mathbf{k}}}{2%
}}\sum_{\lambda =1}^{4}e_{\mathbf{k}\mu }^{\lambda }\{j_{\mathbf{k}%
}\eqslantless \underline{c}_{\mathbf{k}\lambda }(t)\mid e^{i\mathbf{kx}} 
\notag \\
- &\mid &\overline{\underline{c}}_{\mathbf{k}\lambda }(t)\eqslantgtr j_{(-%
\mathbf{k)}}e^{-i\mathbf{kx}}\},  \TCItag{3.8}
\end{eqnarray}%
where $\xi _{\mathbf{p}s},$ $\eta _{\mathbf{p}s}^{+},$ $\eta _{\mathbf{p}s}$
and $\xi _{\mathbf{p}s}^{+}$ are all Grassman numbers,%
\begin{equation*}
\mid \underline{\overline{c}}_{\mathbf{k}\lambda }\eqslantgtr =\left\{ 
\begin{array}{c}
\mid \underline{c}_{\mathbf{k}\lambda }\eqslantgtr ,\lambda =1,2,3, \\ 
-\mid \underline{c}_{\mathbf{k}\lambda }\eqslantgtr ,\lambda =4%
\end{array}%
\right. ,
\end{equation*}%
\begin{equation*}
\mid \overline{c}_{\mathbf{k}\lambda }\eqslantgtr =\left\{ 
\begin{array}{c}
\mid c_{\mathbf{k}\lambda }\eqslantgtr ,\lambda =1,2,3, \\ 
-\mid c_{\mathbf{k}\lambda }\eqslantgtr ,\lambda =4%
\end{array}%
\right. ,
\end{equation*}

\begin{eqnarray}
&\eqslantless &\alpha _{\mathbf{p}s}(t)\mid =\eqslantless \alpha _{\mathbf{p}%
s}\mid e^{-i\omega _{\mathbf{p}}t},\;\mid \alpha _{\mathbf{p}%
s}(t)\eqslantgtr =\mid \alpha _{\mathbf{p}s}\eqslantgtr e^{i\omega _{\mathbf{%
p}}t},\;\omega _{\mathbf{p}}=\sqrt{\mathbf{p}^{2}+m}^{2},  \notag \\
&\eqslantless &\gamma _{\mathbf{k}\lambda }(t)\mid =\eqslantless \gamma _{%
\mathbf{k}\lambda }\mid e^{-i\omega _{\mathbf{k}}t},\;\mid \overline{\mathbf{%
\gamma }}_{\mathbf{k}\lambda }(t)\eqslantgtr =\mid \gamma _{\mathbf{k}%
\lambda }\eqslantgtr e^{i\omega _{\mathbf{k}}t},\;\omega _{\mathbf{k}}=\mid 
\mathbf{k}\mid ,  \TCItag{3.9}
\end{eqnarray}%
where $\alpha =b,a,\underline{a},\underline{b};\gamma =c,\underline{c}.$

We call such operators as $\mid a_{\mathbf{p}s}\rangle \eqslantless 
\underline{a}_{\mathbf{p}s}\mid $ and $\mid \underline{c}_{\mathbf{k}\lambda
}\eqslantgtr j_{\mathbf{k}}$ transformation operators. In the transformation
operators $\mid \alpha \rangle $ and $\langle \alpha \mid $are respectively
a state ket and a state bar which do not change as time $t$, $\eqslantless
\alpha (t)\mid $ and $\mid \overline{\mathbf{\gamma }}(t)\eqslantgtr $ etc.
are operators changing as time $t.$ $\eqslantless \alpha (t)\mid $or $%
\eqslantless \gamma _{\mathbf{k}\lambda }(t)\mid $act on the state$\mid
\alpha \rangle $ or $\mid \gamma _{\mathbf{k}\lambda }\rangle ,$ $\mid
\alpha _{\mathbf{p}s}(t)\eqslantgtr $ or $\mid \overline{\mathbf{\gamma }}_{%
\mathbf{k}\lambda }(t)\eqslantgtr $ act on the state $\langle \alpha \mid $%
or $\langle \gamma _{\mathbf{k}\lambda }\mid .$We call $\mid a_{\mathbf{p}%
s}\rangle $, $\mid b_{\mathbf{p}s}\rangle $ and $\mid c_{\mathbf{k}\lambda
}\rangle $ a F-electron , a F-positron and a F-photon state ket,
respectively, $\mid \underline{a}_{\mathbf{p}s}\rangle $, $\mid \underline{b}%
_{\mathbf{p}s}\rangle $ and $\mid \underline{c}_{\mathbf{k}\lambda }\rangle $
a W-electron, a W-positron and a W-photon state ket, respectively. We can
also name the state bars and the operators in similar method. We call $%
\underline{j}_{\mathbf{k}\text{ }}$and $j_{\mathbf{k}}$ a W-imaginary
current and a F-imaginary current, respectively. It should be pointed out
that $A_{0\mu ,}$ \hspace{0in}( $\hspace{0in}\underline{A}_{0\mu }),$ in
fact, may be written out in W-state and F-operator ( F-state and W-operator)
as well as the form of $\psi _{0}$ ($\underline{\psi }_{0}),$but in this
case, it is not convenient to discuss the Hamiltonian equations.

\subsection{\textit{Properties and multiplication rules of the
transformation operators}.}

We define inner products of the states, products of $j_{\mathbf{k}}$, etc.,
and commutation relations of operator $\mid a_{\mathbf{p}s}\eqslantgtr $
etc. as follows.

\begin{equation}
\langle \alpha _{\mathbf{p}s}\mid \cdot \mid \beta _{\mathbf{p}^{\prime
}s^{\prime }}\rangle =\langle \alpha _{\mathbf{p}s}\mid \beta _{\mathbf{p}%
^{\prime }s^{\prime }}\rangle =\delta _{\alpha \beta }\delta _{\mathbf{pp}%
^{\prime }}\delta _{ss^{^{\prime }}},  \tag{3.10}
\end{equation}

\begin{equation}
\langle \gamma _{\mathbf{k}\lambda }\mid \cdot \mid \gamma _{\mathbf{k}\
^{^{\prime }}\lambda ^{^{\prime }}}^{^{\prime }}\rangle =\left\{ 
\begin{array}{c}
\delta _{\gamma \gamma ^{^{\prime }}}\delta _{\mathbf{kk}^{^{\prime
}}}\delta _{\lambda \lambda ^{^{\prime }}},\lambda =1,2,3, \\ 
-\delta _{\gamma \gamma ^{^{\prime }}}\delta _{\mathbf{kk}^{^{\prime
}}}\delta _{\lambda \lambda ^{^{\prime }}},\lambda =4%
\end{array}%
\right.  \tag{3.11}
\end{equation}

\begin{equation}
\langle \beta _{\mathbf{p}s}\mid \cdot \mid \gamma _{\mathbf{k}\lambda
}\rangle =\langle \gamma _{\mathbf{k}\lambda }\mid \cdot \mid \beta _{%
\mathbf{p}s}\rangle =0,  \tag{3.12}
\end{equation}%
\begin{equation}
\langle 0\mid 0\rangle =1,  \tag{3.13}
\end{equation}

\begin{equation}
\langle 0\mid \alpha \rangle =\langle 0\mid \gamma \eqslantgtr =\langle
\beta _{\mathbf{p}s}\mid \cdot \mid \gamma _{\mathbf{k}\lambda }\rangle =0. 
\tag{3.14}
\end{equation}

\begin{equation}
j_{\mathbf{k}}j_{(-\mathbf{k}^{^{\prime }}\mathbf{)}}=\underline{j}_{\mathbf{%
k}}\underline{j}_{(-\mathbf{k}^{\prime })}=\delta _{\mathbf{kk}^{^{\prime
}}},\text{ \hspace{0in} \hspace{0in} \hspace{0in} \hspace{0in} \hspace{0in} 
\hspace{0in} \hspace{0in} \hspace{0in} \hspace{0in} \hspace{0in} \hspace{0in}
\hspace{0in} \hspace{0in} \hspace{0in} \hspace{0in} }j_{\mathbf{k}}%
\underline{j}_{\mathbf{k}^{\prime }}=0,  \tag{3.15}
\end{equation}%
where $\alpha ,\beta =a,b,\underline{a},\underline{b}$ and $\gamma =c,%
\underline{c}.$Two single-fermion states are anticommutative; two boson
states or a fermion state and a boson state are commutative, i.e.,%
\begin{eqnarray}
&\mid &\alpha _{\mathbf{p}s}\rangle \mid \beta _{\mathbf{p}^{\prime
}s^{\prime }}\rangle =-\mid \beta _{\mathbf{p}^{\prime }s^{\prime }}\rangle
\mid \alpha _{\mathbf{p}s}\rangle ,  \notag \\
Tr &\mid &\alpha _{\mathbf{p}s}\rangle \langle \beta _{\mathbf{p}^{\prime
}s^{\prime }}\mid =-\langle \beta _{\mathbf{p}^{\prime }s^{\prime }}\mid
\alpha _{\mathbf{p}s}\rangle =-\delta _{\alpha \beta }\delta _{\mathbf{pp}%
^{\prime }}\delta _{ss^{\prime }},  \TCItag{3.16}
\end{eqnarray}%
\begin{equation}
\mid \gamma _{\mathbf{k}\lambda }\rangle \mid \gamma _{\mathbf{k}^{\prime
}\lambda ^{\prime }}^{\prime }\rangle =\mid \gamma _{\mathbf{k}^{\prime
}\lambda ^{\prime }}^{\prime }\rangle \mid \gamma _{\mathbf{k}\lambda
}\rangle ,\;Tr\mid \gamma _{\mathbf{k}\lambda }\rangle \langle \gamma _{%
\mathbf{k}^{\prime }\lambda ^{\prime }}^{\prime }\mid =\langle \gamma _{%
\mathbf{k}^{\prime }\lambda ^{\prime }}^{\prime }\mid \gamma _{\mathbf{k}%
\lambda }\rangle =\delta _{\gamma \gamma ^{\prime }}\delta _{\mathbf{kk}%
^{\prime }}\delta _{\lambda \lambda ^{\prime },}  \tag{3.17}
\end{equation}

\begin{equation}
\mid \alpha _{\mathbf{p}s}\rangle \mid \gamma _{\mathbf{k}\lambda }\rangle
=\mid \gamma _{\mathbf{k}\lambda }\rangle \mid \alpha _{\mathbf{p}s}\rangle
,\;Tr\mid \gamma _{\mathbf{k}\lambda }\rangle \langle \alpha _{\mathbf{p}%
s}\mid =\langle \alpha _{\mathbf{p}s}\mid \gamma _{\mathbf{k}\lambda
}\rangle =0.  \tag{3.18}
\end{equation}

The operators $\mid \underline{b}_{\mathbf{p}s}\eqslantgtr $ or $\mid 
\overline{c}_{\mathbf{k}\lambda }\eqslantgtr $ etc. satisfy the same
anticommutation or commutation relations as those for creation and
annihilation operators in the known QED

\begin{equation}
\left\{ \eqslantless \beta _{\mathbf{p}s}(t)\mid ,\mid \beta _{\mathbf{p}%
^{^{\prime }}s^{^{\prime }}}^{^{\prime }}(t)\eqslantgtr \right\} =\delta
_{\beta \beta ^{^{\prime }}}\delta _{\mathbf{pp}^{^{\prime }}}\delta
_{ss^{^{\prime }}},  \tag{3.19}
\end{equation}

\begin{equation}
\left\{ \eqslantless \beta _{\mathbf{p}s}(t)\mid ,\eqslantless \beta _{%
\mathbf{p}^{^{\prime }}s^{^{\prime }}}^{^{\prime }}(t)\mid \right\} =\left\{
\mid \beta _{\mathbf{p}s}(t)\eqslantgtr ,\mid \beta _{\mathbf{p}^{^{\prime
}}s^{^{\prime }}}^{^{\prime }}(t)\eqslantgtr \right\} =0,  \tag{3.20}
\end{equation}

\begin{equation}
\lbrack \eqslantless \gamma _{\mathbf{k}\lambda }(t)\mid ,\mid \gamma _{%
\mathbf{k}\ ^{^{\prime }}\lambda ^{^{\prime }}}^{^{\prime }}(t)\eqslantgtr
]=\left\{ 
\begin{array}{c}
\delta _{\gamma \gamma ^{^{\prime }}}\delta _{\mathbf{kk}^{^{\prime
}}}\delta _{\lambda \lambda ^{^{\prime }}},\lambda =1,2,3, \\ 
-\delta _{\gamma \gamma ^{^{\prime }}}\delta _{\mathbf{kk}^{^{\prime
}}}\delta _{\lambda \lambda ^{^{\prime }}},\lambda =4%
\end{array}%
\right.  \tag{3.21}
\end{equation}%
\begin{eqnarray}
\lbrack &\eqslantless &\gamma _{\mathbf{k}\lambda }(t)\mid ,\eqslantless
\gamma _{\mathbf{k}\ ^{^{\prime }}\lambda ^{^{\prime }}}^{^{\prime }}(t)\mid
]=[\mid \gamma _{\mathbf{k}\lambda }(t)\eqslantgtr ,\mid \gamma _{\mathbf{k}%
\ ^{^{\prime }}\lambda ^{^{\prime }}}^{^{\prime }}(t)\eqslantgtr ]  \notag \\
&=&[\eqslantless \gamma _{\mathbf{k}\lambda }(t)\mid ,\eqslantless \beta _{%
\mathbf{p}s}(t)\mid ]=[\mid \gamma _{\mathbf{k}\lambda }(t)\eqslantgtr ,\mid
\beta _{\mathbf{p}s}(t)\eqslantgtr ]  \notag \\
&=&[\eqslantless \gamma _{\mathbf{k}\lambda }(t)\mid ,\mid \beta _{\mathbf{p}%
s}(t)\eqslantgtr ]=[\eqslantless \beta _{\mathbf{p}s}(t)\mid ,\mid \gamma _{%
\mathbf{k}\ \lambda }(t)\eqslantgtr ]=0,  \TCItag{3.22}
\end{eqnarray}

\begin{equation}
\left[ J,\mid \sigma \rangle \right] =\left[ J,\langle \sigma \mid \right] =%
\left[ J,\mid \sigma (t)\eqslantgtr \right] =\left[ J,\eqslantless \sigma
(t)\mid \right] =0.  \tag{3.23}
\end{equation}%
where $J=j_{\mathbf{k}},$ $\underline{j}_{\mathbf{k}},$ $\sigma =a,$ $b,$ $%
\underline{a},$ $\underline{b},$ $c,$ $\underline{c}.$

We define the action of an operator on a state as follows.

\begin{equation*}
\langle \beta _{\mathbf{p}s}\mid \beta _{\mathbf{p}^{^{\prime }}s^{^{\prime
}}}^{^{\prime }}\eqslantgtr =\langle 0\mid \delta _{\beta \beta ^{^{\prime
}}}\delta _{\mathbf{pp}^{^{\prime }}}\delta _{ss^{^{\prime }}},
\end{equation*}%
\begin{equation}
\eqslantless \beta _{\mathbf{p}s}\mid \cdot \mid \beta _{\mathbf{p}%
^{^{\prime }}s^{^{\prime }}}^{^{\prime }}\rangle =\mid 0\rangle \delta
_{\beta \beta ^{^{\prime }}}\delta _{\mathbf{pp}^{^{\prime }}}\delta
_{ss^{^{\prime }}},,  \tag{3.24}
\end{equation}%
\begin{equation*}
\eqslantless \gamma _{\mathbf{k}\lambda }\mid \gamma _{\mathbf{k}\
^{^{\prime }}\lambda ^{^{\prime }}}^{^{\prime }}\rangle =\left\{ 
\begin{array}{c}
\mid 0\rangle \delta _{\gamma \gamma ^{^{\prime }}}\delta _{\mathbf{kk}%
^{^{\prime }}}\delta _{\lambda \lambda ^{^{\prime }}},\lambda =1,2,3, \\ 
-\mid 0\rangle \delta _{\gamma \gamma ^{^{\prime }}}\delta _{\mathbf{kk}%
^{^{\prime }}}\delta _{\lambda \lambda ^{^{\prime }}},\lambda =4,%
\end{array}%
\right.
\end{equation*}

\begin{equation}
\langle \gamma _{\mathbf{k}\lambda }\mid \gamma _{\mathbf{k}\ ^{^{\prime
}}\lambda ^{^{\prime }}}^{^{\prime }}\eqslantgtr =\left\{ 
\begin{array}{c}
\langle 0\mid \delta _{\gamma \gamma ^{^{\prime }}}\delta _{\mathbf{kk}%
^{^{\prime }}}\delta _{\lambda \lambda ^{^{\prime }}},\lambda =1,2,3, \\ 
-\langle 0\mid \delta _{\gamma \gamma ^{^{\prime }}}\delta _{\mathbf{kk}%
^{^{\prime }}}\delta _{\lambda \lambda ^{^{\prime }}},\lambda =4,%
\end{array}%
\right.  \tag{3.25}
\end{equation}

\begin{eqnarray}
&\eqslantless &\beta _{\mathbf{p}s}\mid \gamma _{\mathbf{k}\ ^{^{\prime
}}\lambda ^{^{\prime }}}^{^{\prime }}\rangle =\eqslantless \gamma _{\mathbf{k%
}\lambda }\mid \beta _{\mathbf{p}s}\rangle =\langle \gamma _{\mathbf{k}%
\lambda }\mid \beta _{\mathbf{p}s}\eqslantgtr =\langle \beta _{\mathbf{p}%
s}\mid \gamma _{\mathbf{k}\ \lambda }\eqslantgtr ,  \notag \\
\langle 0 &\mid &\beta _{\mathbf{p}s}\eqslantgtr =\langle 0\mid \gamma _{%
\mathbf{k}\ \lambda }\eqslantgtr =\langle 0\mid J=J\mid 0\rangle =0. 
\TCItag{3.26}
\end{eqnarray}

The transformation operators have the following properties and observe the
following multiplication rules.

$\left( 1\right) $. Because a transformation operator is a whole, the order
of its two parts cannot exchange, e.g., $\mid \underline{b}_{\mathbf{p}%
s}\eqslantgtr \langle b_{\mathbf{p}s}\mid $ and $\mid \overline{c}_{\mathbf{k%
}\lambda }\eqslantgtr \underline{j}_{\left( -\mathbf{k}\right) }$cannot be
written as $\langle b_{\mathbf{p}s}\mid \mid \underline{b}_{\mathbf{p}%
s}\eqslantgtr $ and $\underline{j}_{\left( -\mathbf{k}\right) }\mid 
\overline{c}_{\mathbf{k}\lambda }\eqslantgtr $, respectively.

$\left( 2\right) .$ Because a transformation operator contains a Grassman
number and a state, when a physics-quantity operator is constructed by the
transformation operators, it is necessary to integrate over the Grassman
numbers and to trace with respect to states in product of the transformation
operators.

According to the rules above we easily obtain the charge operators and the
Hamiltonian operators.

\begin{eqnarray}
Q_{F} &=&\int d^{3}xTr\int \psi _{0}^{+}\psi _{0}  \notag \\
&=&\int Tr\sum_{\mathbf{p}s}\{\mid a_{\mathbf{p}s}(t)\eqslantgtr d\xi _{%
\mathbf{p}s}\langle \underline{a}_{\mathbf{p}s}\mid \underline{a}_{\mathbf{p}%
s}\rangle \xi _{\mathbf{p}s}\eqslantless a_{\mathbf{p}s}(t)\mid  \notag \\
+ &\mid &\underline{b}_{\mathbf{p}s}\rangle \eta _{\mathbf{p}%
s}^{+}\eqslantless b_{\mathbf{p}s}(t)\mid \cdot \mid b_{\mathbf{p}%
s}(t)\eqslantgtr d\eta _{\mathbf{p}s}^{+}\langle \underline{b}_{\mathbf{p}%
s}\mid \}  \notag \\
&=&Tr\sum_{\mathbf{p}s}\{\mid a_{\mathbf{p}s}(t)\eqslantgtr \langle 
\underline{a}_{\mathbf{p}s}\mid \underline{a}_{\mathbf{p}s}\rangle
\eqslantless a_{\mathbf{p}s}(t)\mid  \notag \\
- &\mid &\underline{b}_{\mathbf{p}s}\rangle \eqslantless b_{\mathbf{p}%
s}(t)\mid \cdot \mid b_{\mathbf{p}s}(t)\eqslantgtr \langle \underline{b}_{%
\mathbf{p}s}\mid \}  \notag \\
&=&\sum_{\mathbf{p}s}\left\{ \mid a_{\mathbf{p}s}(t)\eqslantgtr \eqslantless
a_{\mathbf{p}s}(t)\mid -\mid b_{\mathbf{p}s}(t)\eqslantgtr \eqslantless b_{%
\mathbf{p}s}(t)\mid -1\right\}  \TCItag{3.27}
\end{eqnarray}

\begin{eqnarray}
Q_{W} &=&-\int d^{3}xTr\int \underline{\psi }_{0}^{+}\underline{\psi }_{0} 
\notag \\
&=&-\sum_{\mathbf{p}s}\left\{ \mid \underline{b}_{\mathbf{p}s}(t)\eqslantgtr
\eqslantless \underline{b}_{\mathbf{p}s}(t)\mid -\mid \underline{a}_{\mathbf{%
p}s}(t)\eqslantgtr \eqslantless \underline{a}_{\mathbf{p}s}(t)\mid -1\right\}
\TCItag{3.28}
\end{eqnarray}

Similarly to $Q_{F}$ and $Q_{W\text{ }},$ we have%
\begin{eqnarray}
H_{F0} &=&\sum_{\mathbf{p}s}\omega _{\mathbf{p}}[\mid a_{\mathbf{p}%
s}(t)\eqslantgtr \eqslantless a_{\mathbf{p}s}(t)\mid +\mid b_{\mathbf{p}%
s}(t)\eqslantgtr \eqslantless b_{\mathbf{p}s}(t)\mid -1]  \notag \\
+\sum_{\mathbf{k}}\omega _{\mathbf{k}}[\sum_{j=1}^{3} &\mid &c_{\mathbf{k}%
j}(t)\eqslantgtr \eqslantless c_{\mathbf{k}j}(t)\mid -\mid c_{\mathbf{k}%
4}(t)\eqslantgtr \eqslantless c_{\mathbf{k}4}(t)\mid +\frac{1}{2}]. 
\TCItag{3.29}
\end{eqnarray}

\begin{eqnarray}
H_{W0} &=&-\sum_{\mathbf{p}s}\omega _{\mathbf{p}}[\mid \underline{b}_{%
\mathbf{p}s}(t)\eqslantgtr \eqslantless \underline{b}_{\mathbf{p}s}(t)\mid
+\mid \underline{a}_{\mathbf{p}s}(t)\eqslantgtr \eqslantless \underline{a}_{%
\mathbf{p}s}(t)\mid -1]  \notag \\
-\sum_{\mathbf{k}}\omega _{\mathbf{k}}[\sum_{j=1}^{3} &\mid &\underline{c}_{%
\mathbf{k}j}(t)\eqslantgtr \eqslantless \underline{c}_{\mathbf{k}j}(t)\mid
-\mid \underline{c}_{\mathbf{k}4}(t)\eqslantgtr \eqslantless \underline{c}_{%
\mathbf{k}4}(t)\mid +\frac{1}{2}].  \TCItag{3.30}
\end{eqnarray}

It is seen from (3.27)-(3.30) that both energy and charge of the vacuum
state are zero, the energies of the F-states are all positive and the
energies of the W-states are all negative, i.e.,%
\begin{equation}
E_{0}=\langle 0\mid H_{0}\mid 0\rangle =Q_{0}=\langle 0\mid Q\mid 0\rangle
=0,  \tag{3.31}
\end{equation}%
\begin{equation}
\langle F\mid H_{0}\mid F\rangle =\langle F\mid (H_{F0}+\sum_{\mathbf{p}%
s}\omega _{\mathbf{p}}-\frac{1}{2}\sum \omega _{\mathbf{k}})\mid F\rangle >0,
\tag{3.32}
\end{equation}

\begin{equation}
\langle W\mid H_{0}\mid W\rangle =\langle W\mid (H_{W0}-\sum_{\mathbf{p}%
s}\omega _{\mathbf{p}}+\frac{1}{2}\sum \omega _{\mathbf{k}})\mid W\rangle <0.
\tag{3.33}
\end{equation}

It should be pointed out that $E_{0}=0$ and $Q_{0}=0$ do not depend on
existence of negative energies, in fact, provided fields are quantized by
the transformation operators, (3.31) can be obtained$^{[2]}$. (3.33) does
not imply that the masses of the W-particles are negative, oppositely, it is
known from (2.10) that the masses of the W-particles are positive.

The energies and charges of particles can also be written as

\begin{equation}
H_{F0}=\int d^{3}x:[\psi ^{\prime +}\gamma _{4}\left( \gamma _{j}\partial
_{j}+m\right) \psi ^{\prime }+\frac{1}{2}\left( \overset{\cdot }{A^{\prime }}%
_{\mu }\overset{\cdot }{A^{\prime }}_{\mu }+\partial _{j}A_{\nu }^{^{\prime
}}\partial _{j}A_{\nu }^{^{\prime }}\right) ]\ :,  \tag{3.34}
\end{equation}

\begin{equation}
H_{W0}=-\int d^{3}x:[(\underline{\psi }^{^{\prime }+}\gamma _{4}(\gamma
_{j}\partial _{j}+m)\underline{\psi }^{^{\prime }}+\frac{1}{2}(\underline{%
\overset{.}{A}}_{\mu }^{\prime }\underline{\overset{.}{A}}_{\mu }^{\prime
}+\partial _{j}\underline{A}_{\nu }^{\prime }\partial _{j}\underline{A}_{\nu
}^{\prime })]:,  \tag{3.35}
\end{equation}

\begin{equation}
Q=\int d^{3}x:\psi ^{\prime +}\psi ^{\prime }:-\int d^{3}x:\underline{\psi }%
^{^{\prime }+}\underline{\psi }^{^{\prime }}:,  \tag{3.36}
\end{equation}%
where%
\begin{equation}
\psi _{0}^{\prime }\left( x\right) =\frac{1}{\sqrt{V}}\sum_{\mathbf{p}%
s}\left( \eqslantless a_{\mathbf{p}s}(t)\mid u_{\mathbf{p}s}e^{i\mathbf{px}%
}+\mid b_{\mathbf{p}s}(t)\eqslantgtr v_{\mathbf{p}s}e^{-i\mathbf{px}}\right)
,  \tag{3.37}
\end{equation}

\begin{equation}
A_{0\mu }^{\prime }\left( x\right) =\frac{1}{\sqrt{V}}\sum_{\mathbf{k}}\frac{%
1}{\sqrt{2\omega _{\mathbf{k}}}}\sum_{\lambda =1}^{4}e_{\mathbf{k}\mu
}^{\lambda }\left( \eqslantless c_{\mathbf{k}\lambda }(t)\mid e^{i\mathbf{kx}%
}+\mid \overline{c}_{\mathbf{k}\lambda }(t)\eqslantgtr e^{-i\mathbf{kx}%
}\right) ,  \tag{3.38}
\end{equation}

\begin{equation}
\underline{\psi ^{\prime }}_{0}\left( x\right) =\frac{1}{\sqrt{V}}\sum_{%
\mathbf{p}s}\left( \eqslantless \underline{b}_{\mathbf{p}s}(t)\mid u_{%
\mathbf{p}s}e^{i\mathbf{px}}+\mid \underline{a}_{\mathbf{p}s}(t)\eqslantgtr
v_{\mathbf{p}s}e^{-i\mathbf{px}}\right) ,  \tag{3.39}
\end{equation}

\begin{equation}
\underline{A}_{\mu }^{\prime }\left( x\right) =\frac{1}{\sqrt{V}}\sum_{%
\mathbf{k}}\frac{1}{\sqrt{2\omega _{\mathbf{k}}}}\sum_{\lambda =1}^{4}e_{%
\mathbf{k}\mu }^{\lambda }\left( \eqslantless \underline{\overline{c}}_{%
\mathbf{k}\lambda }(t)\mid e^{i\mathbf{kx}}+\mid \underline{c}_{\mathbf{k}%
\lambda }(t)\eqslantgtr e^{-i\mathbf{kx}}\right) ,  \tag{3.40}
\end{equation}

\begin{equation}
\pi _{0\psi }^{\prime }=i\psi _{0\left( x\right) }^{^{\prime }+}=\frac{i}{%
\sqrt{V}}\sum_{\mathbf{p}s}\left( \mid a_{\mathbf{p}s}(t)\eqslantgtr u_{%
\mathbf{p}s}^{+}e^{-i\mathbf{px}}+\eqslantless b_{\mathbf{p}s}(t)\mid v_{%
\mathbf{p}s}^{+}e^{i\mathbf{px}}\right)  \tag{3.41}
\end{equation}

\begin{equation}
\pi _{0\mu }^{\prime }=\dot{A}_{0\mu }^{\prime }\left( x\right) =\frac{-i}{%
\sqrt{V}}\sum_{\mathbf{k}}\sqrt{\frac{\omega _{\mathbf{k}}}{2}}\sum_{\lambda
=1}^{4}e_{\mathbf{k}\mu }^{\lambda }\left( \eqslantless c_{\mathbf{k}\lambda
}(t)\mid e^{i\mathbf{kx}}-\mid \overline{c}_{\mathbf{k}\lambda
}(t)\eqslantgtr e^{-i\mathbf{kx}}\right) ,  \tag{3.42}
\end{equation}

\begin{equation}
\pi _{0\psi }^{\prime }=-i\underline{\psi }_{0}^{^{\prime }+}(x)=\frac{-i}{%
\sqrt{V}}\sum_{\mathbf{p}s}\left( \mid \underline{b}_{\mathbf{p}%
s}(t)\eqslantgtr u_{\mathbf{p}s}^{+}e^{-i\mathbf{px}}+\eqslantless 
\underline{a}_{\mathbf{p}s}(t)\mid v_{\mathbf{p}s}^{+}e^{i\mathbf{px}%
}\right) ,  \tag{3.43}
\end{equation}

\begin{equation}
\underline{\pi }_{0\mu }^{\prime }=-\underline{\dot{A}}_{0\mu }^{\prime }=%
\frac{i}{\sqrt{V}}\sum_{\mathbf{k}}\sqrt{\frac{\omega _{\mathbf{k}}}{2}}%
\sum_{\lambda =1}^{4}e_{\mathbf{k}\mu }^{\lambda }\left( \eqslantless 
\underline{c}_{\mathbf{k}\lambda }(t)\mid e^{i\mathbf{kx}}-\mid \underline{%
\overline{c}}_{\mathbf{k}\lambda }(t)\eqslantgtr e^{-i\mathbf{kx}}\right) , 
\tag{3.44}
\end{equation}%
where the double-dot notation : $\cdots $: is known as normal ordering. An
operator product is in normal ordered form if all operators as $\mid \sigma
\eqslantgtr $ stand to the left of all operators as $\eqslantless \sigma
\mid .$ In contrast with the given QED, (3.31)-(3.33) are the inferences of
the multiplication rules, and are not definition.

\subsection{\textit{Subsidiary condition}}

After the Maxwell field is quantized, the Lorentz condition $\left(
2.18\right) $ is no longer applicable. From $\left( 3.38\right) $ and $%
\left( 3.40\right) $ we have

\begin{equation}
\left( \partial _{\mu }A_{0\mu }^{\prime }\right) ^{+}=\frac{i}{\sqrt{V}}%
\sum_{\mathbf{k}}\frac{\mid \mathbf{k\mid }}{\sqrt{2\omega _{\mathbf{k}}}}%
(\eqslantless c_{\mathbf{k}3}\mid -i\eqslantless c_{\mathbf{k}4}\mid
)e^{ikx},  \tag{3.45}
\end{equation}

\begin{equation}
\left( \partial _{\mu }\underline{A}_{0\mu }^{\prime }\right) ^{-}=\frac{i}{%
\sqrt{V}}\sum_{\mathbf{k}}\frac{\mid \mathbf{k\mid }}{\sqrt{2\omega _{%
\mathbf{k}}}}(\eqslantless \underline{c}_{\mathbf{k}3}\mid -i\eqslantless 
\underline{c}_{\mathbf{k}4}\mid )e^{ikx}.  \tag{3.46}
\end{equation}%
Thus we define the subsidiary condition to be

\begin{equation}
\left( \partial _{\mu }A_{\mu }^{\prime }\right) ^{+}\mid c_{p}\rangle =0, 
\tag{3.47}
\end{equation}

\begin{equation}
\left( \partial _{\mu }\underline{A}_{\mu }^{\prime }\right) ^{-}\mid 
\underline{c}_{p}\rangle =0.  \tag{3.48}
\end{equation}%
$\mid c_{p}\rangle $ and $\mid \underline{c}_{p}\rangle $ are known as
F-physics state ket and W-physics state ket, respectively. From $\left(
3.47\right) -\left( 3.48\right) $ we obtain

\begin{equation}
\mid c_{p}\rangle =\mid c_{T}\rangle \{1+\sum_{\mathbf{k}}f\left( \mathbf{k}%
\right) \mid c_{p\mathbf{k}}\rangle +\cdots +\sum_{\mathbf{k}_{1}\cdot \cdot
\cdot \mathbf{k}_{n}}f\left( \mathbf{k}_{1}\cdots \mathbf{k}_{n}\right) \mid
c_{p\mathbf{k}_{1}}\rangle \cdots \mid c_{p\mathbf{k}_{n}}\rangle +\cdots \},
\tag{3.49}
\end{equation}

\begin{equation}
\mid \underline{c}_{p}\rangle =\mid \underline{c}_{T}\rangle \{1+\sum_{%
\mathbf{k}}\underline{f}\left( \mathbf{k}\right) \mid \underline{c}_{p%
\mathbf{k}}\rangle +\cdots +\sum_{\mathbf{k}_{1}\cdot \cdot \cdot \mathbf{k}%
_{n}}\underline{f}\left( \mathbf{k}_{1}\cdots \mathbf{k}_{n}\right) \mid 
\underline{c}_{p\mathbf{k}_{1}}\rangle \cdots \mid \underline{c}_{p\mathbf{k}%
_{n}}\rangle +\cdots \},  \tag{3.50}
\end{equation}%
where $\mid c_{T}\rangle $ and $\mid \underline{c}_{T}\rangle $ are states
containing only transverse photons, and

\begin{equation}
\mid c_{p\mathbf{k}}\rangle =\mid c_{\mathbf{k}3}\rangle +i\mid c_{\mathbf{k}%
4}\rangle ,  \tag{3.51}
\end{equation}

\begin{equation}
\mid \underline{c}_{p\mathbf{k}}\rangle =\mid \underline{c}_{\mathbf{k}%
3}\rangle +i\mid \underline{c}_{\mathbf{k}4}\rangle .  \tag{3.52}
\end{equation}%
From (2.21), (3.11), (3.22), (3.23) and (3.29)-(3.33) we obtain

\begin{equation}
\langle c_{p\mathbf{k}}\mid c_{p\mathbf{k}^{^{\prime }}}\rangle =\langle 
\underline{c}_{p\mathbf{k}}\mid \underline{c}_{p\mathbf{k}^{^{\prime
}}}\rangle =\langle c_{p\mathbf{k}}\mid H_{0}\mid c_{p\mathbf{k}^{^{\prime
}}}\rangle =\langle \underline{c}_{p\mathbf{k}}\mid H_{0}\mid \underline{c}%
_{p\mathbf{k}^{^{\prime }}}\rangle =0,  \tag{3.53}
\end{equation}

\begin{equation}
\langle c_{p}\mid c_{p^{^{\prime }}}\rangle =\langle c_{T}\mid
c_{T^{^{\prime }}}\rangle ,\text{ }\langle \underline{c}_{p}\mid \underline{c%
}_{p^{^{\prime }}}\rangle =\langle \underline{c}_{T}\mid \underline{c}%
_{T^{^{\prime }}}\rangle ,  \tag{3.54}
\end{equation}

\begin{equation}
\langle c_{p}\mid H_{0}\mid c_{p}\rangle =\langle c_{T}\mid H_{0}\mid
c_{T}\rangle ,\hspace{0pt}  \tag{3.55}
\end{equation}

\begin{equation}
\langle \underline{c}_{p}\mid H_{0}\mid \underline{c}_{p}\rangle =\langle 
\underline{c}_{T}\mid H_{0}\mid \underline{c}_{T}\rangle ,  \tag{3.56}
\end{equation}

\subsection{\textit{Eigenstates of }$H_0.$}

It is easily seen from $\left( 3.16\right) $ and $\left( 3.17\right) $ that
there is at most only one particle in a fermion state denoted by $\mathbf{p}%
s,$ and there may be many particles in a boson state denoted by $\mathbf{k}%
\lambda \mathbf{.}$ Thus, a state in which there are n F-photons or n
W-photons can be represented by

\begin{equation}
\mid n_{\mathbf{k}\lambda }\rangle =\frac{1}{\sqrt{n!}}\mid c_{\mathbf{k}%
\lambda }\rangle ^{n},  \tag{3.57}
\end{equation}

\begin{equation}
\mid \underline{n}_{\mathbf{k}\lambda }\rangle =\frac{1}{\sqrt{n!}}\mid 
\underline{c}_{\mathbf{k}\lambda }\rangle ^{n},  \tag{3.58}
\end{equation}%
From $\left( 3.11\right) $ and $\left( 3.25\right) $ we have

\begin{equation}
\langle n_{\mathbf{k}\lambda }\mid n_{\mathbf{k}\lambda }\rangle =\langle 
\underline{n}_{\mathbf{k}\lambda }\mid \underline{n}_{\mathbf{k}\lambda
}\rangle =\left\{ 
\begin{array}{c}
1,\lambda =1,2,3 \\ 
\left( -1\right) ^{n},\lambda =4%
\end{array}%
\right. ,  \tag{3.59}
\end{equation}

\begin{equation}
\eqslantless c_{\mathbf{k}\lambda }\mid n_{\mathbf{k}\lambda }\rangle
=\left\{ 
\begin{array}{c}
\sqrt{n}\mid (n-1)_{\mathbf{k}\lambda }\rangle ,\lambda =1,2,3 \\ 
-\sqrt{n}\mid (n-1)_{\mathbf{k}\lambda }\rangle ,\lambda =4%
\end{array}%
\right. ,  \tag{3.60}
\end{equation}

\begin{equation}
\eqslantless \underline{c}_{\mathbf{k}\lambda }\mid \underline{n}_{\mathbf{k}%
\lambda }\rangle =\left\{ 
\begin{array}{c}
\sqrt{n}\mid (\underline{n-1})_{\mathbf{k}\lambda }\rangle ,\lambda =1,2,3
\\ 
-\sqrt{n}\mid (\underline{n-1})_{\mathbf{k}\lambda }\rangle ,\lambda =4%
\end{array}%
\right. ,  \tag{3.61}
\end{equation}

\subsection{\textit{The equations of motion}}

From $\left( 3.1\right) $ -$\left( 3.4\right) ,\left( 3.6\right) ,\left(
3.8\right) ,\left( 3.29\right) $ and $\left( 3.30\right) $ , we have

\begin{equation}
i\frac{\partial \psi _{0}}{\partial t}=\left[ \psi _{0},H_{F0}\right] =%
\overset{\wedge }{H}_{0}\psi _{0},  \tag{3.62}
\end{equation}

\begin{equation}
i\frac{\partial \underline{\psi }_{0}}{\partial t}=-[\underline{\psi }%
_{0},H_{W0}]=\overset{\wedge }{H}_{0}\underline{\psi }_{0},  \tag{3.63}
\end{equation}

\begin{equation}
\frac{\partial A_{0\mu }}{\partial t}=-i\left[ A_{0\mu },H_{F0}\right] ,%
\text{ \hspace{0in} \hspace{0in} \hspace{0in} \hspace{0in} \hspace{0in} 
\hspace{0in} \hspace{0in} \hspace{0in} \hspace{0in} }\frac{\partial \dot{A}%
_{0\mu }}{\partial t}=-i[\dot{A}_{0\mu }H_{F0}]=\triangledown ^{2}A_{0\mu },
\tag{3.64}
\end{equation}

\begin{equation}
\frac{\partial \underline{A}_{0\mu }}{\partial t}=i[\underline{A}_{0\mu
},H_{W0}],\text{ \hspace{0in} \hspace{0in} \hspace{0in} \hspace{0in} \hspace{%
0in} }\frac{\partial \underline{\dot{A}}_{0\mu }}{\partial t}=i[\underline{%
\dot{A}}_{0\mu },H_{W0}]=\triangledown ^{2}\underline{A}_{0\mu }.  \tag{3.65}
\end{equation}%
Since $\left[ H_{0,}H_{F0}\right] $ =$\left[ H_{0,}H_{W0}\right] =0$ , $%
H_{F0}$ and $H_{W0}$ are the constants of motion. Thus we have

\begin{equation}
\psi _{0}\left( \mathbf{x,}t\right) =e^{iH_{F0}t}\psi _{0}\left( \mathbf{x,0}%
\right) e^{-iH_{F0}t},  \tag{3.66}
\end{equation}

\begin{equation}
\underline{\psi }_{0}\left( \mathbf{x,}t\right) =e_{0}^{-iH_{W0}t}\underline{%
\psi }_{0}\left( \mathbf{x,0}\right) e^{iH_{W0}t},  \tag{3.67}
\end{equation}

\begin{equation}
A_{0\mu }\left( \mathbf{x,}t\right) =e^{iH_{F0}t}A_{0\mu }\left( \mathbf{x,0}%
\right) e^{-iH_{F0}t},  \tag{3.68}
\end{equation}

\begin{equation}
\underline{A}_{0\mu }\left( \mathbf{x,}t\right) =e^{-iH_{W0}t}\underline{A}%
_{0\mu }\left( \mathbf{x,0}\right) e^{iH_{W0}t},  \tag{3.69}
\end{equation}%
As seen the equations $\left( 3.62\right) -\left( 3.65\right) $ are
consistent with $\left( 2.14\right) -\left( 2.17\right) ,$ respectively.

\section{The physical meanings of that the energy of the vacuum state is
equal to zero.}

It is seen from (3.31) that both energy and charge of the vacuum state are
equal to zero.(3.31) is not relative to the definition for multiplication of
transformation operators. (3.31) holds water provided $H_{0}$ is composed of
transformation operators$^{[2]}$. In contrast with the given QED, (3.31) is
derived without application of the normal ordered product of $H_{0}.$
According to the given QED, before redefining $H_{0}$ as normal-ordered
products $E_{0}\neq 0.$ After redefining $H_{0}$ as normal-ordered products, 
$E_{0}=0.$ But this only transfers the divergence difficulty of the energy
of the ground state. Because we may arbitrarily choose the zero point of
energy in quantum field theory, we can redefine the ground-state energy to
be zero. But in the theory of gravitation, if $E_{0}\neq 0$, $E_{0}$ will
have gravitational effect. Hence we are not at liberty to redefine $E_{0}$
=0 . Thus the knotty problem of the cosmological constant arises in the
given QFT and the relativistic theory of gravitation$^{\left[ 4\right] }$.
In the present theory $E_{0}=0.$ Hence the density of the energy of the
vacuum state $\rho _{vac}=0.$ Thus, from the equation of gravitation field 
\begin{equation*}
R_{\mu \nu }-\frac{1}{2}g_{\mu \nu }R-\lambda g_{\mu \nu }=-8\pi G\left(
T_{\mu \nu }-\rho _{vac}g_{\mu \nu }\right) ,
\end{equation*}%
and data of astronomical observation we can easily determine the
cosmological constant $\lambda .$

We secondly simply discuss the correction originating from $E_{0}=0$ to a
nonperturtational method in quantum field theory.

When one evaluates the energy of a system by a nonperturtational method,
e.g., a Hartree-type approximation$^{\left[ 5\right] },$ it is necessary to
subtract the zero-point energy $E_{0}^{\left[ 6\right] }.$ According to the
given quantum field theory $E_{0}\neq 0,$ while according to the present
theory $E_{0}=0,$ hence we will obtain different results in nature.

We will discuss the two knotty problems above in detail in other papers.

\section{Conclusions}

We suppose that there are both particles with negative energies described by 
$\mathcal{L}_{W}$ and particles with positive energies described by $%
\mathcal{L}_{F\text{ }},$ and $\mathcal{L}_{W}$ and $\mathcal{L}_{F\text{ }}$
are independent of each other and symmetric. From this we present a new
Lagrangian density $\mathcal{L}=\mathcal{L}_{W}$ $+$ $\mathcal{L}_{F\text{ }%
} $and a new quantization method for QED. That the energy of the vacuum
state is equal to zero is naturally obtained. Thus we can easily determine
the cosmological constant according to data of astronomical observation, and
it is possible to correct nonperturbational methods which depend on the
energy of the ground state in quantum field theory.

\section{Acknowledgement}

I am very grateful to professor Zhan-yao Zhao for best support and professor
Zhao-yan Wu for helpful discussions.

\end{document}